# From Monolith to Microservices: A Classification of Refactoring Approaches

Jonas Fritzsch[1], Justus Bogner[2], Alfred Zimmermann[2], Stefan Wagner[1]

[1] Institute of Software Technology, University of Stuttgart, Stuttgart, Germany.
{jonas.fritzsch, stefan.wagner}@informatik.uni-stuttgart.de
[2] Reutlingen University of Applied Sciences, Reutlingen, Germany.
{justus.bogner, alfred.zimmermann}@reutlingen-university.de

**Abstract.** While the recently emerged Microservices architectural style is widely discussed in literature, it is difficult to find clear guidance on the process of refactoring legacy applications. The importance of the topic is underpinned by high costs and effort of a refactoring process which has several other implications, e.g. overall processes (DevOps) and team structure. Software architects facing this challenge are in need of selecting an appropriate strategy and refactoring technique. One of the most discussed aspects in this context is finding the right service granularity to fully leverage the advantages of a Microservices architecture. This study first discusses the notion of architectural refactoring and subsequently compares 10 existing refactoring approaches recently proposed in academic literature. The approaches are classified by the underlying decomposition technique and visually presented in the form of a decision guide for quick reference. The review yielded a variety of strategies to break down a monolithic application into independent services. With one exception, most approaches are only applicable under certain conditions. Further concerns are the significant amount of input data some approaches require as well as limited or prototypical tool support.

**Keywords:** Microservices, Monolith, Modernization, Refactoring, Cloud, Decomposition, Transformation, Modularization, Software Architecture

## 1 Introduction

An increased tendency by organizations to move existing enterprise-scale applications to the cloud can be observed. The reasons to do so are manifold: high availability and redundancy, automatic scaling, easier infrastructure management and compliance with latest security standards ensure a more agile and combined flow of development and operation, also referred to as DevOps [5]. Driven by this new paradigm, the design, build, deployment and maintenance of business applications has fundamentally changed. To overcome this gap and make existing monolithic applications "cloud-ready", they need to run as flexible, loosely-coupled compositions of specialized services, which lately emerged as the Microservices architecture style.

Monolithic applications that have grown over years can become large, complex and in later stages even fossilize [39], meaning the accumulated technical debt results in obscure structures that make the product unmaintainable with a reasonable effort. Even in earlier stages, a single developer or even architect is unable to keep detailed insight into all components and their interfaces. This makes the monolith hard to maintain and cumbersome with regards to adapting newer and better technologies. Furthermore, the effort for changing initial design choices later on requires immense effort. Besides, monolithic applications are often incapable to scale on the module level, but rather per duplicating instances of the whole application. This is in most cases an inefficient approach in responding to quickly changing workloads while maintaining optimal resource utilization.

A new architectural style, referred to as Microservices, promises to address these issues. It started as a trend in software engineering industry practice which was first described in detail by Lewis and Fowler [25]. Contextually related modules have to be identified and encapsulated into a service, providing high cohesion inwards and loose coupling outwards. To leverage most from the design, functionality has to be split up with appropriate granularity. However, building a new application from scratch based on a Microservices architecture can be a very expensive and time-consuming task. On the other hand, the process of refactoring a mature monolithic application into Microservices can be a long-lasting endeavor too, depending on the condition of the system in place.

This study aims to fill the gap in scientific research by comparing and classifying refactoring approaches proposed in academic literature. The results can help architects and developers to gain an overview of currently available refactoring approaches and hereby facilitate their specific transformation process. Researchers may profit from the findings through quickly understanding the current state of the field. The key objective of the study design is formulated as a research question:

**RQ**: What are existing architectural refactoring approaches in the context of decomposing a monolithic application architecture into Microservices and how can they be classified with regards to the techniques and strategies used?

## 2 Architectural Refactoring and Decomposition

Refactoring as an activity to extend the lifetime of existing software products is a behavior preserving code transformation to improve the source code that structurally deteriorated over time [30] or accumulated technical debt [39]. According to Pirkelbauer [33], agile software development methodologies benefit in particular due to frequent changes. Plenty of research has been conducted in this area already, which mainly targets refactoring at source code level. Fowler et al. consolidated the field in their well-known book "Improving the design of Existing Code, more than 70 Patterns explained" [15]. Dietrich distinguishes code-level from architectural refactoring by referring to the latter ones as high-impact refactorings [11]. They can be seen as architectural activities that remove a particular architectural smell while improving one or more quality attributes, without changing the system's scope and functionality [41]. Moreover, it may result in an altered organizational structure [35], which is an interesting aspect: According to Conway's Law [10], organizations tend to produce system designs that reflect the organization's communication structures [23, 38]. Consequently, architecture and organization are interdependent to some degree, which furthermore distinguishes the process from pure code refactoring. Drivers for a refactoring are feature extensions and design changes [33], but also anti-patterns [8] and code smells [15], whereas such high-impact refactorings are rather driven by requirements to run software in the cloud (platform changes, deployment and release cycle changes) as well as interconnected organizational changes. In contrast to code-level refactoring, architectural refactoring is common in the context of adopting Microservices.

From a software architects' perspective, a proper decomposition into services with the appropriate granularity can be seen as the main challenge in a refactoring process: In general, one could imagine various ways to split a system into smaller parts. Amundsen [2] outlines a few of them, e.g. based on *implementation technology* (computationally heavy services written in C may be separated from chatty components using Node.js) or based on *geography* (also specific legal, commercial or cultural aspects). Besides them, one could think of even other viewpoints, like the architectural style, certain non-functional requirements, personal experiences or education. The characteristics of Microservices promote following the functional decomposition perspective [37]. In this context it is referred to as decomposition around business capabilities. Dependencies throughout the technical layers are hereby greatly reduced, whereas a rather lightweight integration layer on top is a common solution to integrate the resulting Microservices [26].

So, what are the means to identify business capabilities in a monolith? Lewis and Fowler [25] bring the notion of a *bounded context* into effect. It originates from Evans book Domain-Driven Design [13], which provides the means to identify such contexts within a complex domain [25]. According to Richardson, bounded contexts can be separated through decomposing by verbs (use cases) or by nouns (resources) [36]. Newman stresses the term *seam* from Michael Feathers book "Working Effectively with Legacy Code" [14]. It similarly describes a way to separate portions of code that can be treated independently from other parts and hereby obtain "loosely coupled and strongly cohesive" [29] Microservices. In practice, the lack of a universally valid algorithm that guides the decomposition process makes it to "somewhat of an art", as Richardson points out [74]. Extracting a domain model from an application's code base can be a significant challenge. If incorrectly applied, it can lead to architectures that combine the drawbacks of both styles, Monolith and Microservices.

## 3   Related Work

Our literature review has revealed a lack of systematic guidance on the refactoring process for existing monolithic applications. Several publications discussing Microservices also cover the aspect of migrating monoliths to Microservice-based architectures to some extent [22, 25, 29, 36], but the topic is still evolving. A systematic mapping study conducted in 2016 identified 3 out of 21 studies dealing with migration topics [31], while Di Francesco et al. found 16 out of 71 migration-related studies during their review in 2017 [16]. The papers found were mainly solution proposals, followed by experience reports and opinion papers. The field is not mature yet, Microservices migration and architectural refactoring are still referred to as future trends [31]. The very recent and comprehensive study by Balalaie et al. compiles a set of empirically identified design patterns for Microservices migration and rearchitecting [3]. The patterns originate from observations of medium to large-scale industrial projects. Compared to our work, the concepts are presented on a higher level of abstraction and do not cover specifics of concrete approaches proposed in literature. Still, the study complements our work in terms of empirical values. Widening the scope to Service-based Systems in general, there is a mature state of research regarding Service-Oriented Architecture (SOA). According to Bogner et al. [6], Microservices and SOA "share a large set of design-related commonalities". Klose et. al. for instance discuss the identification of services for SOA development from a business point of view [21]. Although the suitability for Microservices may be limited due to the differences of the architectural styles, the included comparison of approaches regarding service identification mark a decent overview at that time. To the best of our knowledge, there is currently no holistic literature review of refactoring approaches and decomposition techniques available that facilitates this process. Our study attempts to fill this research gap.

## 4   Research Method and Search Strategy

By means of a literature review, existing refactoring techniques in the Microservices context are identified, investigated, classified and presented in textual and visual form. Brereton et al. propose a three-step review process that serves as a basic structure for this review: planning, conducting, and documenting [7]. Fundamental constraints of a literature review are the databases to query and the search strings to use. For the used queries, three of the most frequented scientific libraries and indexing systems in computer science have been selected: ACM Digital Library, IEEE Xplore and Google Scholar. The choice of these databases and indexing systems is guided by the fact that they have been proven most relevant for conducting systematic literature reviews in the software engineering field [32]. Other aspects are their high accessibility and ability to export search results conveniently. Fig. 1 illustrates the basic steps for our literature search.

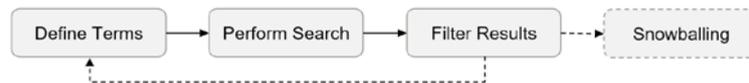

**Fig. 1.** Search strategy used for the Review.

The following search string(s) have been used for querying the databases:

```
("microservice" OR "micro-service") [AND "monolith*"]
[AND ("refactor" OR "transform" OR "migrat*" OR
 "decompos*" OR "partition*" OR "granular*")]
```

The obtained studies have been filtered according to a set of selection criteria: Only peer reviewed articles published in English have been included, the abstract had to clearly show a contribution towards the research question and we expected a documented validation of proposed approach. Guidelines recommend to use a snowballing activity applied on the list resulting from the initial selection [40]. The initial search results yielded by the queries have been enlarged by such a snowballing activity. Finally, a qualitative assessment of the studies has been performed by focusing on technical depth, recency and relevance of the content presented.

## 5   Results

The performed literature review identified a variety of studies with different orientation, coverage and level of detail. Many of them were tailored to specific scenarios, focusing on specific requirements or aspects while not discussing the theoretical background. Ten approaches provided an adequate level of abstraction and potential for generalization according to the underlying strategy used to steers the decomposition (see Table 1). The work by Chen et al. [9] was published after completion of the review and thus did not go into the list of selected publications.

**Table 1.** Reviewed Publications.

| | List of Authors and Publications |
|---|---|
| 1 | Escobar, D. et al.: Towards the understanding and evolution of monolithic applications as microservices. In: Proceedings of 42nd Latin American Computing Conference, CLEI. (2016) [12] |
| 2 | Levcovitz, A. et al.: Towards a Technique for Extracting Microservices from Monolithic Enterprise Systems. In: 3rd Brazilian Workshop on Software Visualization, Evolution and Maintenance (VEM). pp. 97–104 (2015) [24] |
| 3 | Ahmadvand, M., Ibrahim, A.: Requirements reconciliation for scalable and secure microservice (de)composition. In: Proceedings - 2016 IEEE 24th International Requirements Engineering Conference Workshops, REW 2016. pp. 68–73 (2016) [1] |
| 4 | Baresi, L. et al.: Microservices Identification Through Interface Analysis. In: ESOCC 2017: Service-Oriented and Cloud Computing. pp. 19–33 (2017) [4] |
| 5 | Gysel, M. et al.: Service cutter: A systematic approach to service decomposition. In: Lecture Notes in Computer Science. pp. 185–200 (2016) [17] |
| 6 | Mazlami, G. et al.: Extraction of Microservices from Monolithic Software Architectures. In: 2017 IEEE International Conference on Web Services (ICWS). pp. 524–531 (2017) [27] |
| 7 | Mustafa, O., Gómez, J.M.: Optimizing economics of microservices by planning for granularity level Experience Report. (2017) [28] |
| 8 | Hassan, S. et al.: Microservice Ambients: An Architectural Meta-Modelling Approach for Microservice Granularity. In: Proceedings - 2017 IEEE International Conference on Software Architecture, ICSA. pp. 1–10 (2017) [18] |
| 9 | Klock, S. et al.: Workload-Based Clustering of Coherent Feature Sets in Microservice Architectures. Proc. - 2017 IEEE Int. Conf. Softw. Archit. ICSA. 11–20 (2017) [20] |
| 10 | Procaccianti, G. et al.: Towards a MicroServices Architecture for Clouds. VU University Amsterdam (2016) [34] |

### 5.1 Classification

While analyzing the selected approaches, we identified distinct decomposition strategies. They determine the required artefacts (besides source code) as an input, the granularity of the resulting services and if the approach can be applied to greenfield-developments in addition. Out of the reviewed studies, the following categories have been defined by grouping similar strategies:

- *Static Code Analysis aided* approaches require the application's source code and derive a decomposition from it (through possible intermediate stages).
- *Meta-Data aided* approaches require more abstract input data, like architectural descriptions in form of UML diagrams, use cases, interfaces or historical VCS data.
- *Workload-Data aided* approaches aim to find suitable service cuts by measuring the application's operational data (e.g. communication, performance) on module or function level and use this data to determine a fitting decomposition and granularity.
- *Dynamic Microservice Composition* approaches try to solve the problem more holistically by describing a Microservices runtime environment. Other than the above categories, the resulting set of services is permanently changing in each iteration of re-calculating the best-fitting composition (based on e.g. workload).

Table 2 and Table 3 give an overview of the reviewed approaches. The classification defined above can be found in the *Type* column. The *Applicability* column distinguishes between approaches that support Microservices greenfield developments and others that focus on existing monolithic applications. Other constraints like technology-restrictions are listed in this column as well. *Strategy* points out the utilized decomposition strategy. *Atomic Unit, Granularity* indicates the smallest unit that the approach is able to handle, which in the end determines the possible range of granularity. Some approaches automatically calculate the granularity, i.e. number of resulting services, whereas others leave it up to the user. *Input* and *Output* list artefacts needed and produced by the approach. Some approaches describe metrics for a result evaluation, which can be found under *Result Evaluation*. Four of the approaches offer tool-support, as the respective column shows. Our review revealed a general lack in this area, which is mandatory to achieve a certain degree of automation. It hinders an empirical evaluation and thorough assessment of the approaches. Lastly, the column *Validation* shows the kind of method used to validate the approach like experiments, case studies or proof-of-concepts.

**Table 2.** Overview of Decomposition Approaches, Part 1.

| # | Approach | Authors (Year) | Type | Applicability | Strategy | Atomic Unit, Granularity |
|---|---|---|---|---|---|---|
| 1 | Towards the understanding and evolution of monolithic applications as microservices | Escobar, et. al. (2016) [12] | SCA, based on Static Code Analysis from Java annotations | MO, JEE multi-tier applications | calculate clusters of EJBs that form a microservice, identify data types through Java annotations | atomic unit: EJB, adjustable granularity during clustering threshold provided by user |
| 2 | Towards a Technique for Extracting Microservices from Monolithic Enterprise Systems | Levcovitz, et. al. (2016) [24] | SCA, focusing on multi-tier applications | MO, multi-tier applications consisting of at least 3 tiers | construct microservice candidates based on dependencies between facades and database tables, bridged by business functions | atomic unit: set of facades, business functions, database table, granularity as result |
| 3 | Requirements reconciliation for scalable and secure microservice (de)composition | Ahmadvand, et. al. (2016) [1] | MDA, focusing on Security and Scalability | GR+MO, application defined by use cases and requirements | calculate microservice decomposition based on security and scalability requirements | atomic unit as defined in use case diagrams |
| 4 | Microservices Identification Through Interface Analysis | Baresi, et. al. (2017) [4] | MDA, based on semantic similarity of (Open)API specification | GR+MO | calculate suitable service cuts through clustering of interface specifications according to their semantic similarity | single operation as provided by OpenAPI spec., granularity parameterizable |
| 5 | Service Cutter: A systematic approach to service decomposition | Gysel, et. al. (2016) [17] | MDA, extracts coupling information from software engineering artifacts (ERM, use cases) | GR+MO | calculate clustering of nanoentities to form microservices based on number of weighted properties, clustering algorithm exchangeable | nanoentity (data, operation or artifact), granularity as result or input param, depending on algorithm |
| 6 | Extraction of Microservices from Monolithic Software Architectures | Mazlami, et. al. (2017) [27] | MDA, based on Version Control Meta Data | MO, applications having meaningful VCS meta data | calculate decomposition via graph-based clustering out of version history by either: Logical, Semantic or Contributor Coupling | class as atomic unit, granularity as result |
| 7 | GranMicro: A Black-Box Based Approach for Optimizing Microservices Based App's | Mustafa, et. al. (2017) [28] | WDA, black box-based approach, considering non-functional requirements | MO, web-applications generating expressive access logs | utilize web usage mining techniques to optimize service decomposition based on non-functional requirements | functional units that can be identified through web access logs |
| 8 | Microservice Ambients: An Architectural Meta-Modelling Approach for Microservice Granularity | Hassan, et. al. (2017) [18] | DMC, dynamic composition, model granularity at runtime | GR+MO | define architectural elements (Ambients) with adaptable boundaries, use workload data for adaptation of granularity at runtime | "Unit of mobility" as abstract definition of an atomic unit |
| 9 | Workload-based Clustering of Coherent Feature Sets in Microservice Architectures | Klock, et. al. (2017) [20] | DMC, Dynamic composition approach for workload-optimized deployment | GR+MO | calculate optimal deployment and granularity based on workload using a genetic algorithm | feature as atomic unit (chunk of functionality that delivers business value) |
| 10 | Towards a MicroServices Architecture for Clouds | Procaccianti, et. al. (2016) [34] | DMC, MDA, data-driven, bottom-up approach | GR+MO | bottom-up, data-driven, process-mining algorithm | functional property, granularity adapts dynamically |

**Table 3.** Overview of Decomposition Approaches, Part 2.

| # | Input | Output | Result Evaluation | Tool Support | Validation |
|---|---|---|---|---|---|
| 1 | Source Code (Java) | visualization in four different diagrams: EJB data, EJB shared Types, MS, MS invocation | metrics based on source code | n/a | JEE application with 74.566 LoC, 624 classes and 35993 methods |
| 2 | Source Code | candidate list of microservices | n/a | n/a | case study on a 750 KLOC banking application |
| 3 | use cases (UML) with assessment of security and scalability requirements | candidate list of microservices | n/a, announced for future research | n/a | sample application |
| 4 | OpenAPI specification of interface; reference vocabulary (as fitness function) | candidate list of microservices | qualitative, no metrics | experimental prototype of decomposition tool and sample datasets, https://github.com/mgarriga/decomposer | 452 OpenAPI specifications, comparison of samples with results from 5 SW-engineers, comparison with Service Cutter (#4) |
| 5 | Domain Model (ERM) and User Representations (Use Cases, characteristics of nano-entities and roles) in JSON | candidate list of microservices, export to JSON, graphical representation of service and dependencies | qualitative service design checklist assessing service cut (Excellent, Expected, Acceptable, Unreasonable) | Service Cutter, open source prototype implementing the approach, https://github.com/ServiceCutter/ServiceCutter | case studies: fictitious trading system and DDD sample application "Cargo Tracking", performance tests |
| 6 | Source Code and VCS meta data | candidate list of microservices | quality of service cut using custom metrics: Team Size Reduction (tsr), average domain redundancy (adr) | POC available as open source Java project https://github.com/gmazlami/microserviceExtraction-backend (and -frontend) | experiment using a set of sample code bases from open-source projects (200 to 25000 commits, 1.000 to 500.000 LOC, 5 to 200 contributors) |
| 7 | Web access logs | diagram of service model | performance metrics (Response Time, CPU utilization) | n/a | sample e-bookshop web application |
| 8 | aspect-oriented description of the software architecture using the Ambient-PRISMA Textual Language | microservice composition with dynamic granularity adaptation at runtime, based on predefined parameters indicating QoS | qualitative evaluation on Effectiveness/ Expressiveness of modelling and Facilitating design time and runtime analysis | n/a | experiment using a hypothetical application for an online movie subscription-based system |
| 9 | representation of the architecture by a set of features, workload model | descriptive and visual output of suggested model, resulting in concrete MS architectures at runtime | performance metrics measuring the quality of a deployment | MicADO (Microservice Architecture Deployment Optimizer) URL: see publication | case study using ERP software "AFAS" (27 features with a total of 238 properties and 72 dependency relations between features) |
| 10 | properties or blocks extracted from source code, capabilities (non-functional) | microservice composition | n/a | n/a | proof of concept: sample application for "synthetic video processing" |

**Table 4:** Legend to Tables 2 and 3.

| | Type |
|---|---|
| **SCA** | Static Code Analysis aided (either Source Code or more abstract artefacts like architectural UML diagrams or APIs of the applications architecture) |
| **MDA** | Meta-Data aided (version control history data, non-functional requirements) |
| **WDA** | Workload-Data aided (gathered during runtime, like performance data or web-access logs) |
| **DMC** | Dynamic Microservice Composition (approach to model or adapt Service composition/ granularity at runtime based on workload data) |
| | **Applicability** |
| **GR** | Microservices-Greenfield development |
| **MO** | Monolith-Migrations |
| **GR+MO** | applicable for both scenarios |

### 5.2 Decision Guide

Fig. 2 illustrates the essentials of the presented approaches in form of a decision guide. The architect planning to migrate a monolithic application to Microservices can use this flow chart to quickly find the appropriate approach for a specific scenario. Starting on top, a set of alternatives will lead to the most appropriate approach first, symbolized by the number. Should this option not fulfill the architect's requirements, the dashed line will lead back to the main thread and propose the next best alternative. Each approach is labeled with its associated type (symbolized by the orange ellipse), according to its classification (column *Type*). Should all approaches be discarded, the last one proposed will be "Service Cutter" with No. 5, at the bottom right of the flow chart. It can be seen as a general-purpose approach offering the most mature tool support as of date of this review. However, the approach requires a comprehensive specification of the system including coupling criteria, which may not always be available to such extent [4].

Approaches treating granularity as a dynamically changing factor are grouped in a single box and not further differentiated. These approaches describe a Microservices runtime environment in contrast to a fixed partitioning determined at design-time. As such environments are not discussed in necessary detail here, the condensed depiction will account for their complexity.

# Decision Guide

## for Decomposition Approaches of Monolithic Applications

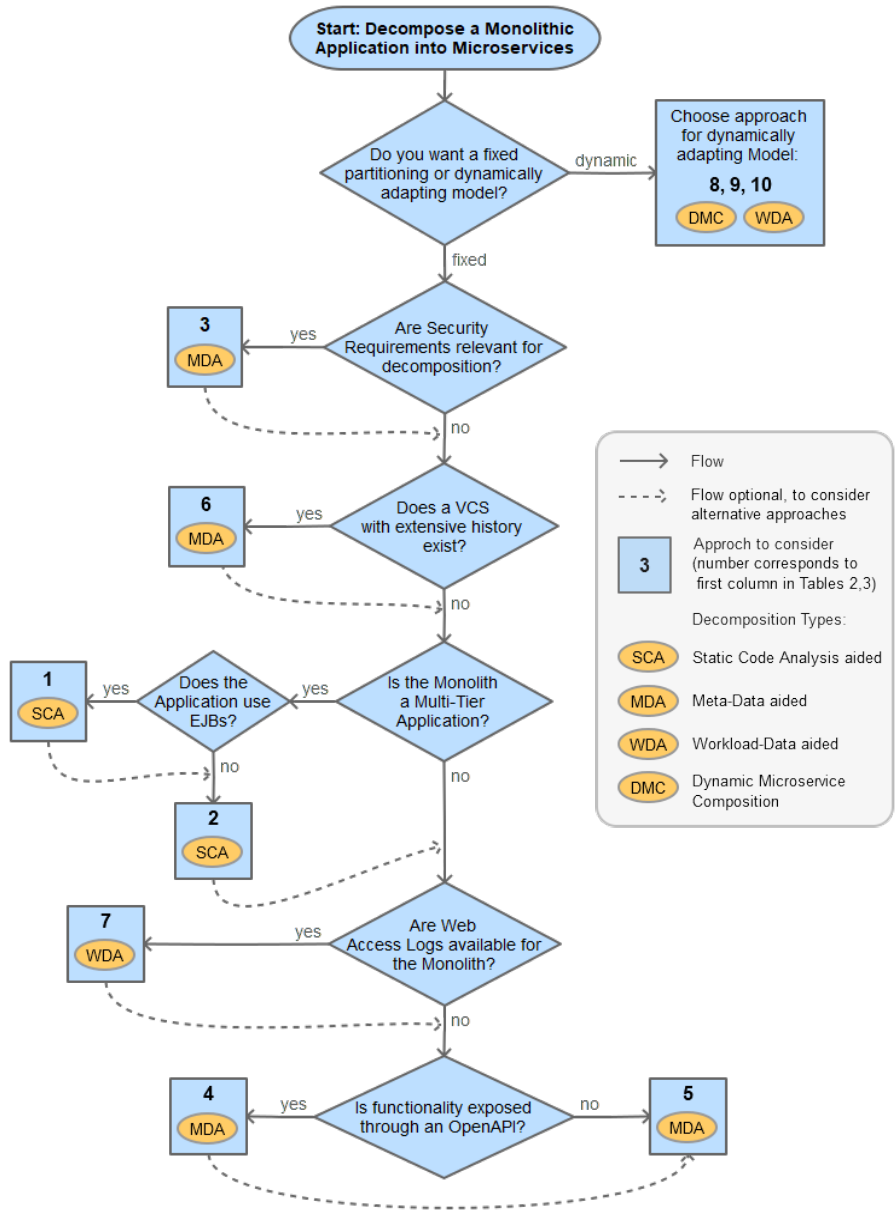

**Fig. 2.** Decision Guide for Decomposition Approaches.

# 6     Conclusion

By means of a literature review we identified and categorized 10 recently proposed architectural refactoring approaches for transforming monolithic applications into Microservices. The approaches have been categorized into four groups by the underlying strategy used for the decomposition, which can be seen as the most challenging step from a software architect's perspective. Thereby we answer our initially phrased research question.

In general, the findings reveal a shortage of practically applicable approaches that offer adequate tool support and metrics to verify the results. Almost all of the reviewed approaches are not universally applicable and require different sets of input data. Thus, an accompanying decision guide in form of a flow chart has been created to help readers in quickly identifying the appropriate approach for a certain scenario. The most structured and universal method has been proposed by Gysel et al. [17], which can be seen as a solid basis for further research. However, the practical applicability is limited due to its dependence on a "detailed and exhaustive specification of the system" [4]. Microservices architecture as a field "rooted in practice" [16] is widely discussed in industry. It can be expected that further research will very likely reveal new approaches that can be incorporated and thus extend the findings of this study. Potential future research could focus on testing different approaches using an adequate example or real-world application. To do so, quality attributes and related metrics to assess the quality of a decomposition should be defined in a first step.

Several threats to validity have to be mentioned for this research. The conducted review did not follow the guidelines of a systematic literature review as proposed by e.g. Kitchenham and Charters [19], which would improve repeatability and reproducibility of the results and thus guarantee appropriate scientific rigor. For the systematic classification and presentation of the results Petersen et. al [32] provide a set of guidelines accordingly. The candidates for this review were obtained from only three academic search engines. Furthermore, the selected refactoring techniques have been investigated only theoretically. Thus, all results stem from assertions of the authors or other publications. A thorough investigation and assessment would require to exercise and test the approaches on the basis of one or more sample applications, better yet, real world systems. The decision guide has been created to suggest or rule out certain approaches for specific environments or indicate the limited applicability in this respect. However, it has neither been systematically constructed nor validated by architects. Future research on the topic of Microservices migration may consider these points to achieve more precise results.

Our future work in this field will focus on (1) novel approaches that combine static code analysis with operations data generated during runtime to achieve an optimally tailored partitioning, (2) quality attributes and related metrics to quantitatively assess the result of a decomposition in advance and (3) other means to automate and facilitate the transformation of monolithic architectures out of large, heterogeneous code bases.